\documentclass[11pt,onecolumn,draftclsnofoot]{IEEEtran}
\usepackage{amsfonts}
\IEEEoverridecommandlockouts

\ifCLASSINFOpdf
\else
\fi
\usepackage{psfig}
\usepackage{epsf}
\usepackage{bm} 
\usepackage{slashbox}
\usepackage{booktabs}
\usepackage{pifont}
\usepackage{times,amsmath,color,amssymb,graphicx,geometry, epsfig,psfrag,subfigure,algorithm}
\usepackage{enumerate}

\begin{document}
\title{State of the Art, Taxonomy, and Open Issues on Cognitive Radio Networks with NOMA}
\author{\IEEEauthorblockN{Fuhui Zhou, \emph{Member, IEEE}, Yongpeng Wu, \emph{Senior Member, IEEE},\\ Ying-Chang Liang, \emph{Fellow, IEEE}, Zan Li, \emph{Senior Member, IEEE},\\Yuhao Wang, \emph{Senior Member, IEEE}, Kai-Kit Wong, \emph{Fellow, IEEE}}
\thanks{Fuhui Zhou and Yuhao Wang are with the School of Information Engineering, Nanchang University, Nanchang, 330031, China,  (e-mail: zhoufuhui@ncu.edu.cn, wangyuhao@ncu.edu.cn).

Yongpeng Wu is with Shanghai Key Laboratory of Navigation and Location Based Services, Shanghai Jiao Tong University, Minhang, 200240, China (Email:yongpeng.wu2016@gmail.com).

Ying-Chang Liang is with University of Sydney, Sydney, NSW 2006, Australia, and University of Electronic Science and Technology of China, Chengdu, China (liangyc@ieee.org).

Zan Li is with the Integrated Service Networks Lab of Xidian University, Xi’an, 710071, China (e-mail: zanli@xidian.edu.cn).

Kai-Kit Wong is with the Department of Electronic and Electrical Engineering, University College London, WC1E 7JE, United Kingdom (e-mail: kai-kit.wong@ucl.ac.uk).
}
\thanks{TThe research was supported by the National Natural Science Foundation of China (61701214, 61701301, 61661028, 61631015, 61561034, 61761030 and 61501356) The Young Natural Science Foundation of Jiangxi Province (20171BAB212002), the China Postdoctoral Science Foundation (2017M610400), and The Postdoctoral Science Foundation of Jiangxi Province (2017KY04).}}
\maketitle
\begin{abstract}
The explosive growth of mobile devices and the rapid increase of wideband wireless services call for advanced communication techniques that can achieve high spectral efficiency and meet the massive connectivity requirement. Cognitive radio (CR) and non-orthogonal multiple access (NOMA) are envisioned to be important solutions for the fifth generation wireless networks. Integrating NOMA techniques into CR networks (CRNs) has the tremendous potential to improve spectral efficiency and increase the system capacity. However, there are many technical challenges due to the severe interference caused by using NOMA. Many efforts have been made to facilitate the application of NOMA into CRNs and to investigate the performance of CRNs with NOMA. This article aims to survey the latest research results along this direction. A taxonomy is devised to categorize the literature based on operation paradigms, enabling techniques, design objectives and optimization characteristics. Moreover, the key challenges are outlined to provide guidelines for the domain researchers and designers to realize CRNs with NOMA. Finally, the open issues are discussed.
\end{abstract}
\vbox{} 
\IEEEpeerreviewmaketitle
\section{Introduction}
\IEEEPARstart{T}{he} explosive growth of mobile devices, the rapidly increasing demand on the broadband and high-rate communication services, such as augmented reality (AR) and virtual reality (VR), and the fixed spectrum assignment policy result in the increasingly severe spectrum scarcity problem. According to the third Generation Partnership Project (3GPP), compared with the fourth generation (4G) networks, the fifth generation (5G) networks are required to achieve 1000 times higher system capacity, 10 times higher spectral efficiency (SE), and 100 times higher connectivity density \cite{J. G. Andrews}. Moreover, among these requirements, meeting the system capacity is the most important but probably the most challenging one due to the limited spectrum resource. Thus, it is imperative to develop advanced communication techniques that can achieve high SE as well as massive wireless connectivity. As a promising technique, cognitive radio (CR) has drawn significant attention in both industry and academia due to its high SE \cite{S. Haykin}. It can enable the secondary network (or called the unlicensed network) to access the licensed frequency bands of the primary network by using adaptive transmission strategies while protecting the quality-of-service (QoS) of the primary one.

Besides CR, non-orthogonal multiple access (NOMA) techniques are promising to improve SE and user connectivity density \cite{Z. Ding}, \cite{L. Dai}. Unlike the conventional orthogonal multiple access (OMA) techniques, NOMA techniques allow multiple users simultaneously access the network at the same time and the same frequency band by using non-orthogonal resources, such as different power levels or low-density spreading codes. In \cite{L. Dai}, the authors have classified the existing dominant NOMA schemes into two categories based on the non-orthogonality resources, namely, power-domain NOMA and code-domain NOMA. The pros and cons of different NOMA schemes have been comprehensively discussed and the applied situations have been characterized. The non-orthogonality enables NOMA techniques to have advantages in SE, massive connectivity, and low transmission latency at the cost of the mutual interference and the implementation complexity of the receiver \cite{L. Dai}, \cite{S. M. R. Islam}. In order to decrease the mutual interference among users, multi-user detection techniques (e.g.,  successive interference cancellation (SIC) for power-domain NOMA) are required. However, the implementation complexity of the receiver structure increases with the number of NOMA users. User-pairing techniques are efficient to decrease the complexity of the receiver structure by pairing users into multiple clusters \cite{S. M. R. Islam}. The number of users in a cluster can be managed. Users in the same cluster can enjoy different services by using NOMA and different clusters can coexist by using OMA \cite{S. M. R. Islam}.

It is envisioned that integrating NOMA techniques into CR networks (CRNs) has the tremendous potential to improve SE and the number of users to be served \cite{Z. Yang}-\cite{Y. Liu}. Recently, the authors in \cite{Z. Yang}-\cite{Y. Liu} have demonstrated that CRNs with NOMA can achieve significant SE gains compared with CRNs with OMA. Moreover, in \cite{L. Lv}-\cite{F. Kader}, it was shown that NOMA techniques are useful for the secondary user (SU) to efficiently cooperate with the primary user (PU) and obtain the spectrum access opportunity. The performances of the SU and the PU can be simultaneously improved by designing an appropriate cooperation mechanism in CRNs with NOMA \cite{L. Lv}-\cite{F. Kader}. Furthermore, Y. Zhang \emph{et al.} in \cite{Y. Zhang} have shown that the energy efficiency (EE) of CRNs with NOMA can be higher than that of CRNs with OMA. However, to make it widely adoptable, there are several challenges required to be tackled. If these challenges are not well understood and appropriately addressed, the SE of CRNs may be decreased, or even the designed CRNs cannot work. For example, the mutual interference between the primary network and the secondary network may be severer due to the non-orthogonal nature of NOMA, which can decrease the SE of CRNs \cite{Y. Liu}. Although CR and NOMA have been extensively and individually studied, there is not much relevant work focused on CRNs with NOMA. The research on integrating NOMA into CRNs is in its infancy. Thus, it is of great importance to understand the challenges and the benefits of CRNs with NOMA. To this end, we conduct the study.

The contributions of our work are summarized as follows. Firstly, we investigate, categorize and review the state-of-the-art research efforts made in the domain of CRNs with NOMA. Secondly, a taxonomy is devised based on the conducted survey considering operation paradigms, enabling techniques, design objectives and optimization characteristics. Finally, we clarify the challenges and discuss the open issues as the future research directions.

The rest of the article is organized as follows. Section II presents the state-of-the-art studies on CRNs with NOMA. Then, in Section III, we discuss the devised taxonomy of CRNs with NOMA. The challenges and the open research issues are discussed in Section IV. Finally, Section V concludes the paper.
\section{State of The Art}
\begin{figure*}[!ht]
\centering
\includegraphics[width=0.8\textwidth]{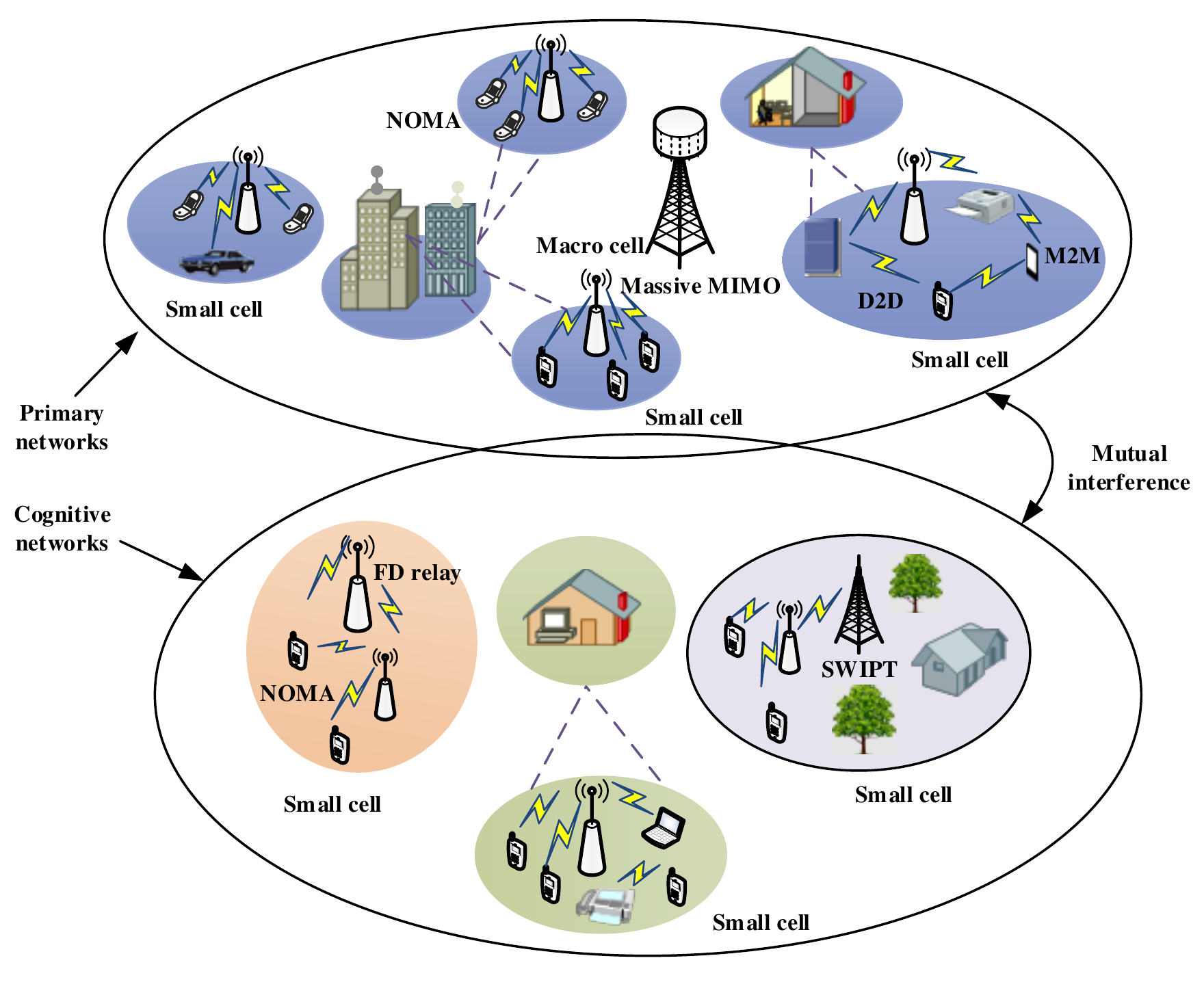}
\caption {\space\space Illustration of CRNs with NOMA.}
\label{system model}
\end{figure*}
In this section, we present the state-of-the-art research efforts made in the domain of CRNs with NOMA based on the considered communication type and operation paradigm. Fig. 1 provides an illustration of CRNs with NOMA. It is envisioned that multiple advanced communication techniques and different multiple access techniques will be exploited in CRNs with NOMA in order to cater to diverse service requirements, different applications, and cells with different scales, and also to efficiently manage the mutual interference between the primary network and the secondary network. Specifically, massive multiple-input multiple-output (MIMO) and full-duplex (FD) communication can be applied for achieving high SE and EE; wireless charging and simultaneous wireless information and power transfer (SWIPT) are good choices for enjoying long-time services; device-to-device (D2D) and machine-to-machine (M2M) communications are appropriate for achieving low latency and high capacity. Fig. 2 presents an overview of the state-of-the-art investigations on CRNs with NOMA. The research has been conducted for about three years, and an increasing attention is drawn.
\begin{figure*}[!ht]
\centering
\includegraphics[width=0.8\textwidth]{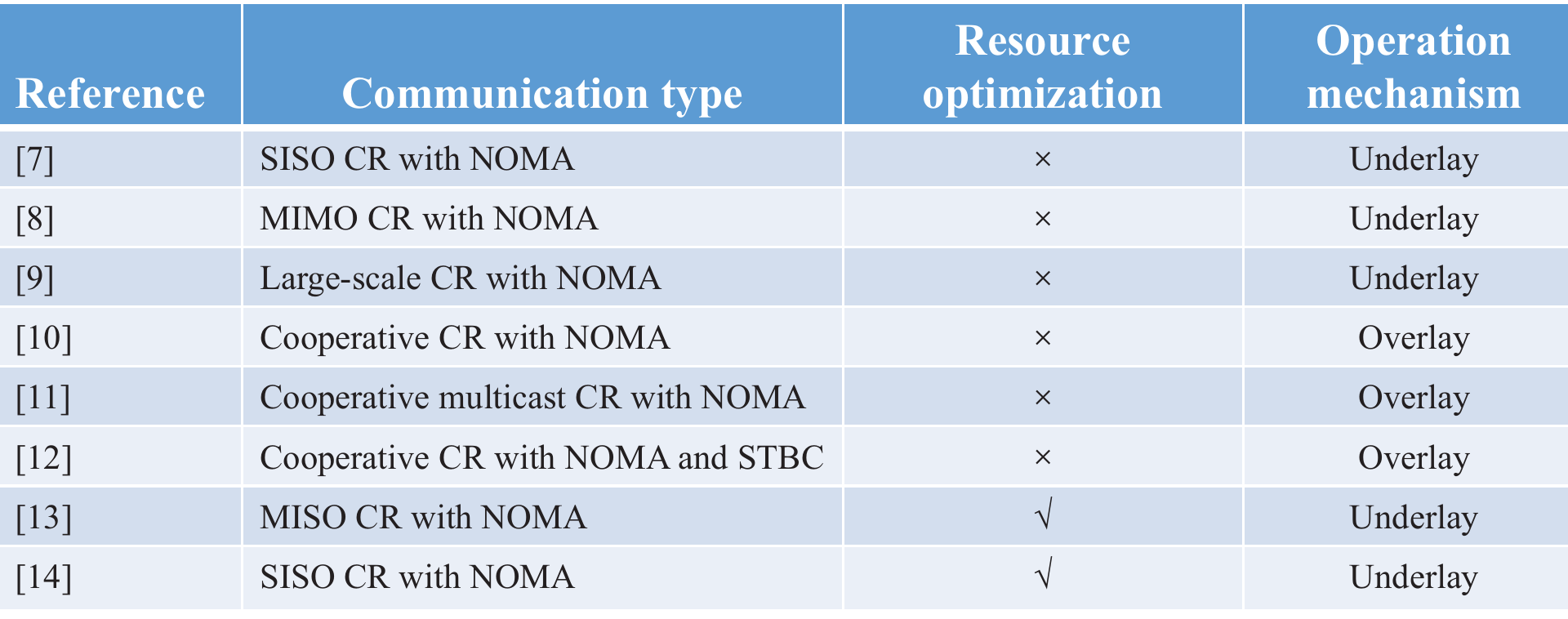}
\caption {\space\space Comparison of communication techniques and operation paradigms used in CRNs with NOMA.}
\label{system model}
\end{figure*}

\begin{figure*}[!ht]
\centering
\includegraphics[width=0.8\textwidth]{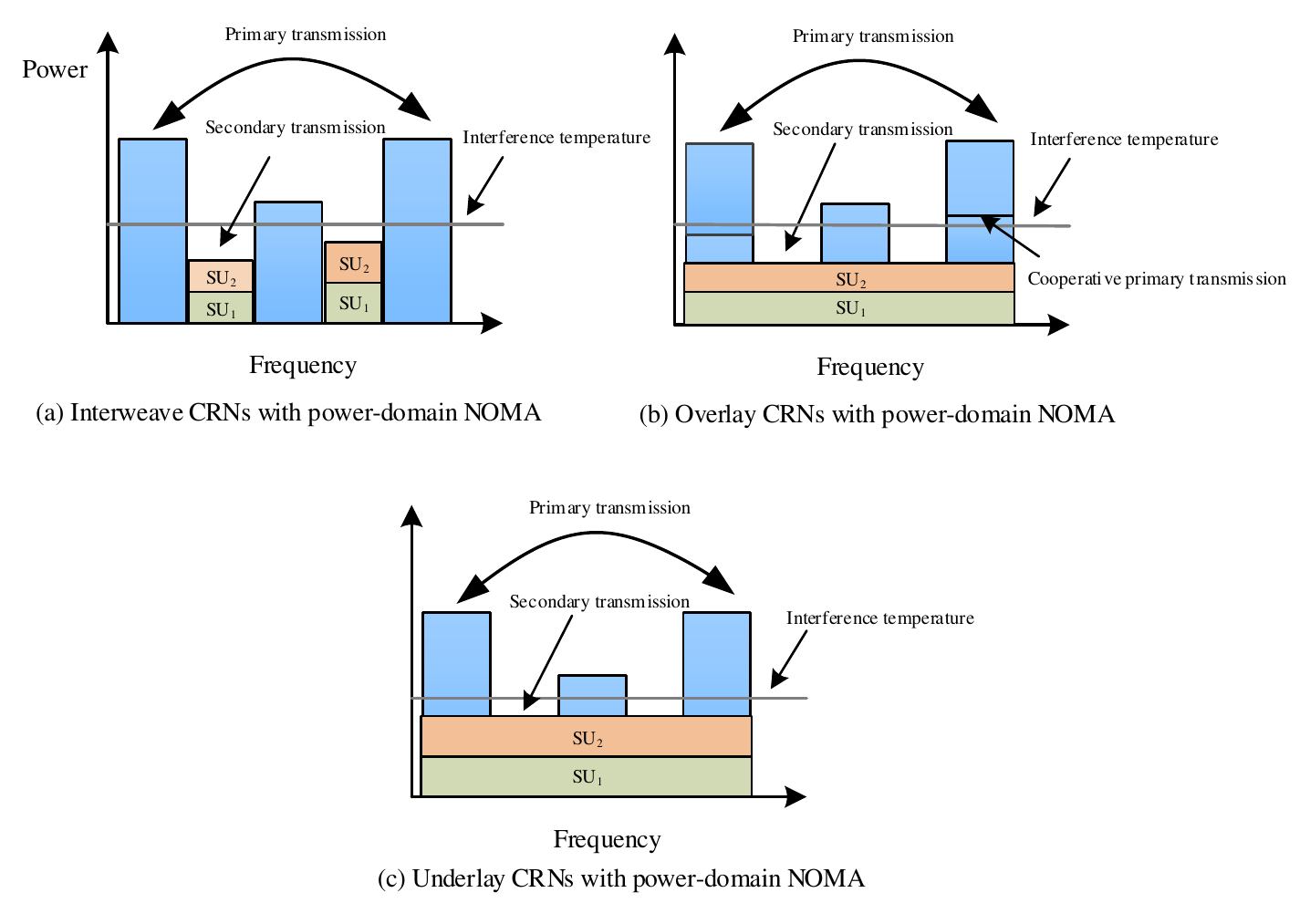}
\caption {\space\space CRNs with power-domain NOMA: three operation paradigms.}
\label{system model}
\end{figure*}

In CRNs, there are three operation paradigms, namely, the interweave mode, the overlay mode and the underlay mode \cite{A. Goldsmith}. Fig. 3 illustrates three operation paradigms in CRNs with power-domain NOMA, where the secondary network serves for multiple secondary users (SUs) with different power levels. In the interweave mode, the frame structure consists of a sensing slot and a data transmission slot \cite{A. Goldsmith}. During the sensing slot, spectrum sensing is performed to detect whether the frequency bands are occupied by the primary users (PUs) or not. During the data transmission slot, the secondary network can access the frequency bands of PUs and provide services for multiple SUs by using NOMA when PUs are detected to be inactive, otherwise it continues performing spectrum sensing in order to find available frequency bands. Since spectrum sensing is imperfect in practice due to the shadowing and fading \cite{S. Haykin}, a false decision that PUs are inactive can be made. Thus, an interference constraint (e.g., the interference temperature is tolerable) is required to protect the QoS of PUs. Regarding the overlay mode, the secondary network provides a cooperation for the primary network in order to obtain the chance to access the frequency band. In the underlay mode, the secondary network coexists with the primary network on the condition that the interference caused by SUs is tolerable to PUs. The frame structure under the underlay and overlay mode only consists of a data transmission slot since these two networks can coexist at the same frequency band and the same time \cite{S. Haykin}. Up to now, most of the researches on CRNs with NOMA are conducted under the underlay mode (e.g., \cite{Z. Yang}-\cite{Y. Liu}) and the overlay mode (e.g., \cite{L. Lv}-\cite{F. Kader}). To the authors' best knowledge, there is no work on CRNs with NOMA under the interweave mode.

In \cite{Z. Yang}, the authors first introduced power-domain NOMA techniques into the underlay CRNs. Under the constraint on guaranteeing the QoS the PU, the SU coexists with the PU by using power-domain NOMA where the SU and the PU transmit information at the same frequency band and the same time with different power levels. Moreover, the authors have investigated the impact of user pairing on the performance of CRNs with power-domain NOMA. It was shown that CRNs can achieve a good performance by using the user-pairing strategy that two users with the best channel conditions are selected to pair. It is different from the pairing strategy in the conventional wireless networks with power-domain NOMA that select the user with the best channel condition and the user with the worst channel condition to pair.

To enhance the performance gains of NOMA, the application of MIMO techniques to CRNs with power-domain NOMA was investigated under the underlay mode \cite{Z. Yang1}. The impacts of the power allocation strategy based on CR and user pairing on the performance were analyzed. It was shown that the performance of CRNs with NOMA can be greatly improved by using MIMO techniques. In \cite{Y. Liu}, the authors have studied the performance achieved by using different power allocation strategies in large-scale underlay CRNs with NOMA. It obtained the conclusion that NOMA can outperform the conventional OMA in underlay CRNs when the target data rate and the power allocation coefficients were properly designed. The works in \cite{Z. Yang}-\cite{Y. Liu} have established the basis for the future research related to the application of power-domain NOMA into underlay CRNs.

Under the overlay mode, \cite{L. Lv}-\cite{F. Kader} have analyzed the performance of the application of NOMA in different CRNs. In \cite{L. Lv}, L. Lv \emph{et al.} designed a cooperative transmission scheme based on NOMA for overlay CRNs. The aim of the proposed scheme was to allow the SU to access the frequency band of the PU and simulatively provide a cooperation for the PU. Moreover, the authors have qualitatively analyzed how NOMA techniques can be beneficial to both the PU and the SU. Furthermore, simulation results have revealed that the overlay CRNs can achieve performance gains by using NOMA compared with that by using the conventional OMA. The authors in \cite{L. Lv2} extended the work to the multicast CRNs and proposed a dynamic cooperative scheme under the overlay mode. A new metric was defined to evaluate the cooperative benefit. It was shown that the performances of both the PU and the SU can be significantly improved by using the proposed cooperative scheme. In \cite{F. Kader}, a two-phase cooperative protocol was proposed based on the Alamouti space time block coded NOMA in overlay CRNs. The efficiency of the proposed protocol was verified compared with the conventional scheme based on the superpositing coding with respect to the outage probability and the ergodic capacity.
\subsection{Resource Optimization in CRNs with NOMA}
The works in \cite{Z. Yang}-\cite{F. Kader} mainly focused on analyzing the achievable performance based on the proposed strategy or the proposed protocol in CRNs with NOMA. Besides the performance analysis, resource optimization is also of great importance. An optimal resource allocation strategy not only can improve the performance of SUs but also can better protect the QoS of the PU \cite{S. Haykin}. The design of an optimal resource allocation strategy depends on the optimization objective and constraints of CRNs with NOMA.

In \cite{Y. Zhang}, the authors proposed an efficient algorithm to optimize the EE of underlay CRNs with NOMA based on the sequential convex approximation method. The considered CRNs with NOMA are general networks which consider an arbitrary number of PUs. The simulation results have shown that NOMA can achieve EE gains in underlay CRNs compared with OMA. The authors in \cite{M. Zeng} designed an optimal power allocation strategy for the underlay CRNs with NOMA. Moreover, the characteristic of NOMA was exploited to design the optimal power allocation algorithm. Furthermore, the superiority of the proposed algorithm was verified by simulation results. To the best of authors' knowledge, the research on the resource optimization for the CRNs with NOMA is very limit. There is much work to do for the application of NOMA techniques into CRNs.
\section{Taxonomy}
\begin{figure*}[!ht]
\centering
\includegraphics[width=0.8\textwidth]{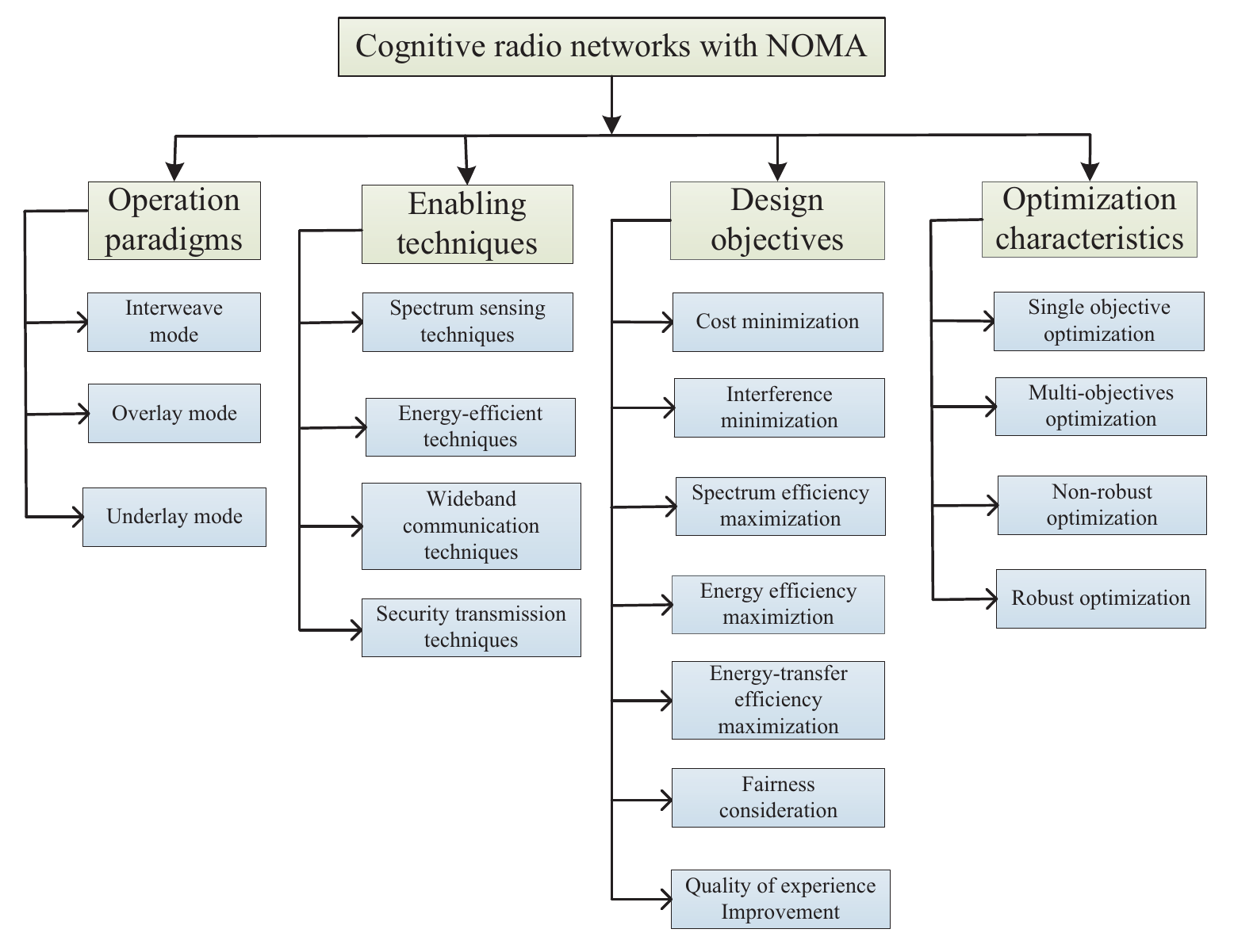}
\caption {\space\space Taxonomy of CRNs with NOMA.}
\label{system model}
\end{figure*}
By summarizing the works in \cite{Z. Yang}-\cite{M. Zeng}, Fig. 4 is given to show the taxonomy of CRNs with NOMA. The taxonomy proposed in this article is based on the parameters, namely, the operation paradigms, the enabling techniques, design objectives and the optimization characteristics.
\subsection{Operation Paradigms}
The operation paradigm of CRNs with NOMA clarifies the spectrum sharing mode between the primary network and the secondary network. The selection of the operation paradigm depends on the implementation complexity of CRNs, the QoS requirement of PUs, and the cooperative ability between these two networks \cite{S. Haykin}. For example, the interweave mode is appropriate for the frequency bands allocated to the very high frequency/ultra high frequency (VHF/UHF) television (TV) service since SUs can access these bands only when they are not occupied \cite{A. Goldsmith}; the overlay mode can be selected when SUs can provide cooperation for PUs (e.g., SUs with strong channel conditions transmit PUs' information or transfer energy to PUs by using NOMA techniques in order to obtain the spectrum access opportunity) \cite{L. Lv}; The underlay mode is a sound choice when the tolerable interference is  high \cite{Z. Yang}. No matter which mode is selected, the QoS of the PU should be protected \cite{A. Goldsmith}.

Compared with the conventional CRNs with OMA, CRNs with NOMA have their own particularities. Specifically, CRNs with NOMA serve for multiple SUs by using non-orthogonal resources and multi-users detection techniques are required in the SUs' receiver in order to decrease the mutual interference. Moreover, under the interweave mode, spectrum sensing algorithms are required to detect the non-orthogonal signals when the primary network transmits information to PUs by using NOMA. They are different from spectrum sensing algorithms for OMA signals due to the correlation among the samples of the received signals. Under the overlay mode, CRNs with NOMA provide cooperation for PUs by superimposing PUs' signals onto SUs's signals \cite{L. Lv}-\cite{F. Kader}. Under the underlay mode, the resource allocation strategy of CRNs with NOMA is different from that of CRNs with OMA. For example, when the power-domain NOMA is applied, a high power level is allocated for the SU with poor channel conditions in order that the SU with good channel conditions can decode the information transmitted to the worse-case SU and cancel the interference caused by the worse-case SU \cite{Z. Yang}-\cite{Y. Liu}.
\subsection{Enabling Techniques}
CRNs with NOMA leverage various techniques to efficiently manage interference and improve the EE, SE and the security \cite{L. Dai}, \cite{S. M. R. Islam}. These techniques include spectrum sensing techniques, energy-efficient techniques, wideband communication techniques and secure transmission techniques. Spectrum sensing techniques are required in the interweave CRNs with NOMA to find the available frequency band. Energy-efficient techniques aim for improving EE during the transmission process. In order to satisfy the high capacity demand of SUs, wideband communication techniques are required. In addition, CRNs with NOMA also use secure transmission techniques to guarantee the privacy of the transmitted information.

In the interweave mode, spectrum sensing techniques with high performance are beneficial for the interference management \cite{S. Haykin}. Up to now, many spectrum sensing algorithms have been proposed, such as energy detection, eigenvalue-based detection, cyclostationary-based detection, etc. \cite{S. Haykin}. Although these algorithms can be applied when NOMA techniques are used in the primary network, the traditional performance analysis based on the independence assumption that the samples of the received signal related to PUs are independent is invalid due to the exploitation of the non-orthogonal resources. For example, the detection probability of energy detection that is derived based on the independence and Gaussian distribution assumption is invalid when PUs's signals are NOMA signals \cite{S. Haykin}. Moreover, how to exploit the non-orthogonal characteristic to improve performances of these algorithms or develop novel spectrum sensing algorithms are still open issues.

Energy-efficient techniques can be divided into two categories. One is that these techniques operate in an energy-efficient way, such as massive MIMO and cooperative relay \cite{Z. Yang1}. The other is that these techniques can harvest energy from nature or the surrounding electromagnetic radiation, such as SWIPT \cite{E. Boshkovska}. In the former, the energy-efficient operation is from the space diversity and cooperation diversity. For example, when massive MIMO is applied in CRNs with NOMA, the cognitive base station (CBS) equips with a large number of antennas (e.g., more than 100) can serve for multiple SUs by using NOMA and its transmitted power can be low so that the interference caused to PUs is tolerable \cite{Z. Yang}. Moreover, the EE can be further improved by designing energy-efficient precoding matrixes for different SUs \cite{Y. Zhang}.

Regarding to energy harvesting (EH) techniques, there are two mechanisms, namely, the wireless charging mechanism and the SWIPT mechanism \cite{S. M. R. Islam}. By using the wireless charging mechanism, the CBS provides energy supply for NOMA SUs in the downlink and receives information from NOMA SUs in the uplink. When the second mechanism is applied, the CBS can simulatively transmit information and transfer energy to NOMA SUs in the downlink; and can simultaneously harvest energy and receive information from NOMA SUs in the uplink. The structures of the CBS and SUs are relatively simple in the first mechanism whereas a higher hardware implementation complexity at the CBS and SUs is required in the second mechanism \cite{E. Boshkovska}. To practically realize SWIPT, the received signal needs to be split into two parts, one for harvesting energy and one for decoding information. When the SWIPT mechanism is chosen in CRNs with NOMA, the hardware implementation complexity of the CBS and NOMA is very high since the multi-user detection techniques and signal splitting techniques are required. In order to decrease the complexity, there are three protocols to realize signal splitting, namely, the time-domain protocol, the power-domain protocol and the antenna-domain protocol. These protocols are designed based on the splitting domain. For example, in the time-domain protocol, the CBS and SUs switch in time between energy harvesting and information decoding.

Wideband communication techniques are expected to significantly enhance the capacity of CRNs with NOMA since a wide band is available \cite{Z. Ding}. There are mainly two wideband communication techniques, namely, millimeter-wave (mmW) communications and multiband communications. MmW communications operate on mmWave bands between 30 and 300 GHz. It is envisioned that integrating mmW communications into CRNs with NOMA can provide a ultra-wide bands services and allow massive connectivity of different devices with diverse service requirements \cite{Z. Ding}. For the multiband communications, it provides the potential to efficiently manage the interference between the primary network and the secondary network, obtain better channel maintenance by reducing handoff frequency and improve the flexility of system design since multiple frequency bands can be accessed. For example, a hybrid multiple access mechanism that combines the orthogonal frequency division multiple access (OFDMA) with NOMA can be designed to decrease the mutual interference between these two networks and improve the capacity of SUs \cite{S. M. R. Islam}. Specifically, SUs can be paired into several clusters. Each cluster uses NOMA to serve for its SUs and different clusters work on different frequency bands.

Due to the broadcast nature of NOMA and the open nature of CR, malicious NOMA SUs may exist and illegitimately access PUs' frequency bands or change the radio environment \cite{E. Boshkovska}. As a result, the legitimate SU is unable to use frequency bands of the PU or has his confidential transmitted information intercepted. Thus, secure transmission techniques are of crucial importance in CRNs with NOMA. The traditionally cryptographic technique depends on secret keys to guarantee the communication confidentiality but increases the computational and communication overhead. It is not the optimal selection in CRNs with NOMA due to the limited resource. Alteratively, physical-layer security is a good choice to improve the security of CRNs with NOMA since it exploits the physical characteristics (e.g., multipath fading, propagation delay, etc.) of wireless channels to achieve secure communications \cite{E. Boshkovska}.
\subsection{Objectives}
CRNs with NOMA can be deployed to improve the number of users in different situations. Based on the requirements and the available resource, the design of CRNs with NOMA has different objectives. The key objectives are cost minimization, interference minimization, SE maximization, EE maximization, energy-transfer efficiency maximization, fairness, and the experience (QoE) of SUs \cite{S. M. R. Islam}, \cite{E. Boshkovska}. The cost minimization is recognized to minimize the transmission power of the CBS when it provides services for multiple SUs. The interference minimization focuses on minimizing the interference caused to PUs while considering the QoS of the SU. The objective of the SE maximization devotes to optimizing the capacity of SUs on the condition that the QoS of the PU is protected. EE is an important metric in the design of the future CRNs, which is defined as the ratio of the total capacity to the total power consumed at the CBS \cite{Y. Zhang}. The EE maximization focuses on maximizing the EE of CRNs with NOMA while the SE is satisfied and the interference caused to PUs is tolerable. When energy harvesting techniques are applied in CRNs with NOMA, the energy-transfer efficiency maximization is the focus, which is defined as the ratio of the harvested power to the consumed power \cite{E. Boshkovska}.

The objective of fairness tries to provide fairness among SUs while considering the QoS of the PU. Although NOMA can provide a good fairness among users compared with OMA \cite{Z. Ding},  the fairness among SUs in CRNs with NOMA is also required to be considered. Up to now, several fairness criterions have been proposed, such as the max-min fairness, the proportional fairness and the harmonic fairness \cite{S. M. R. Islam}. Specifically, the max-min fairness can guarantee the performance of the SU with the worst channel condition; the proportional fairness is applied to balance user fairness and network sum-rate; and the harmonic fairness is achieved by maximizing the harmonic mean of all users' rate \cite{S. M. R. Islam}. The improvement of QoE is to enhance the overall acceptability of an application or service from the perspective of SUs \cite{E. Boshkovska}. It is generally evaluated by
using a mean opinion score \cite{E. Boshkovska}. In order to realize the prescribed objective, resource allocation and optimization techniques are of crucial importance.
\subsection{Optimization Characteristic}
The optimization characteristic classifies the design requirement of CRNs with NOMA and the characteristic of the design objective. The single objective optimization is appropriate in CRNs with NOMA where the system design only emphasizes optimizing one performance metric (e.g., SE, EE, etc.) while other metrics can be treated as constraints of the optimization problem. In CRNs with NOMA, due to the limited resource, there exist multiple tradeoffs among various optimization objectives, such as the tradeoff between the SE and the EE, the tradeoff between the SE and the energy-transfer efficiency \cite{E. Boshkovska}. However, it is difficult to achieve a good tradeoff by using the single objective optimization since it over-emphasizes the importance of one metric. Alternatively, the multi-objective optimization can provide a good tradeoff among various conflicting objectives. Thus, it is applicable in CRNs with NOMA where multiple objectives are required to be jointly optimized. Compared with the optimization in CRNs with OMA, the optimization in CRNs with NOMA is more challenging due to the exploitation of the non-orthogonal resource. Particularly, the constraints for the feasibility of multi-user detection should be imposed \cite{Y. Zhang}, \cite{M. Zeng}. For example, the signal-to-interference-plus noise ratio (SINR) constraints are required to successfully implement SIC under the predefined decoding order when the power-domain NOMA is applied \cite{Y. Zhang}.

The non-robust optimization represents the optimization problem formulated under the assumption that all the channel state information (CSI) can be perfectly obtained in CRNs with NOMA \cite{M. Zeng}. Although this assumption is impractical, investigations on the non-robust optimization are meaningful since the results obtained under the non-robust optimization can present the theoretical limit analysis for the design of CRNs with NOMA. A robust design can be realized by using the robust optimization techniques \cite{E. Boshkovska}. It considers the practical case that the perfect CSI in CRNs with NOMA cannot be obtained due to the presence of quantization errors, time delay and the limited feedback resource. The robust optimization in CRNs with NOMA is different from that in CRNs with OMA. The impact of the imperfect CSI on multi-user detection should be considered. For example, robust optimization in CRNs with power-domain NOMA needs to consider imperfect SIC that the interference signals from other SUs cannot be perfectly removed, which increases the difficulty to solve the robust optimization problem.
\section{Challenges and Future Directions}
\begin{figure*}[!ht]
\centering
\includegraphics[width=0.8\textwidth]{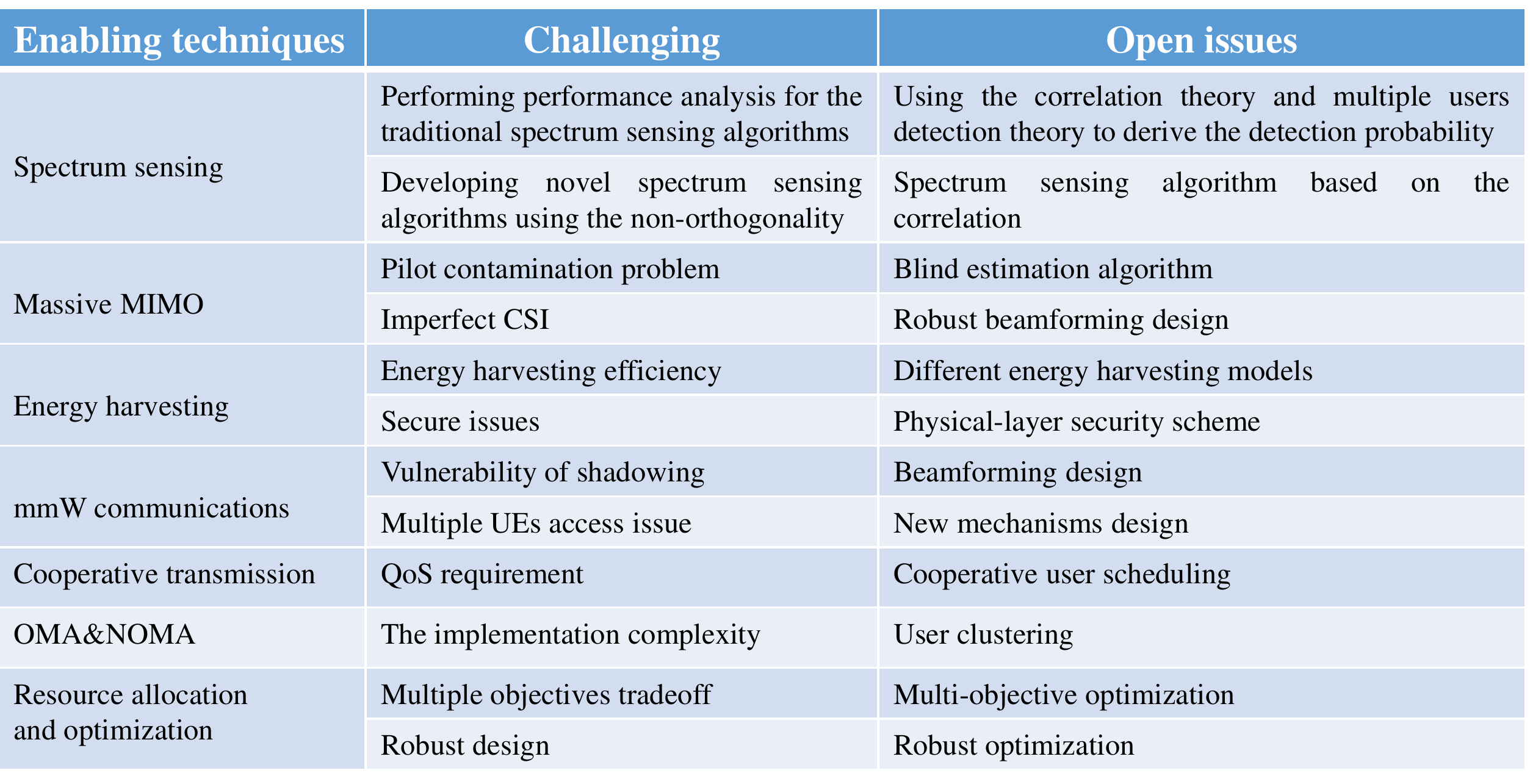}
\caption {\space\space Challenging and open issues in CRNs with NOMA.}
\label{system model}
\end{figure*}
In this section, the challenges and open issues are highlighted when the enabling techniques are implemented in CRNs with NOMA. Fig. 5 summarizes the challenges and open issues.

\textbf{Spectrum sensing:} In the interweave mode, spectrum sensing with a high performance is beneficial for the interference management in CRNs with NOMA. When NOMA techniques are not applied in the primary network, the traditional spectrum sensing algorithm (e.g. energy detection, eigenvalue-based spectrum sensing, etc.) can be exploited and the performance analysis is valid. When NOMA techniques are exploited to provide services for multiple PUs, although the traditional spectrum sensing algorithm can also work, the performance analysis based on the independence assumption that the samples of the received signal are independent is invalid. The reason is that the correlation among different samples exists due to the non-orthogonal resources. It is challenging to derive the probability of detection when the correlation exists among samples. Up to now, how to use the correlation theory and multiple users detection theory to derive the probability of detection is an open issue. On the other hand, how to use the non-orthogonality to develop novel spectrum sensing algorithms is an interesting research issue.

\textbf{Massive MIMO:} Massive MIMO can significantly improve SE and efficiently manage the interference of CRNs with NOMA \cite{Z. Yang1}. However, there are two challenges to be addressed. The pilot contamination problem that the estimation of the channels between the CBS and one NOMA SU is contaminated by the channels between that CBS and other NOMA SUs may be severer due to the non-orthogonal resources \cite{S. M. R. Islam}. Thus, it is important to design blind estimation algorithms to overcome this problem in CRNs with NOMA. Moreover, a large amount of CSI is required to design the optimal precoding scheme that maximizes the capacity of the SU or the EE of the secondary network. however, it is difficult to obtain perfect CSI in practice due to the limited resource and the existence of quantization errors \cite{E. Boshkovska}. How to design robust precoding schemes considering imperfect multi-user detection is a further open issue.

\begin{figure*}[!ht]
\centering
\includegraphics[width=0.8\textwidth]{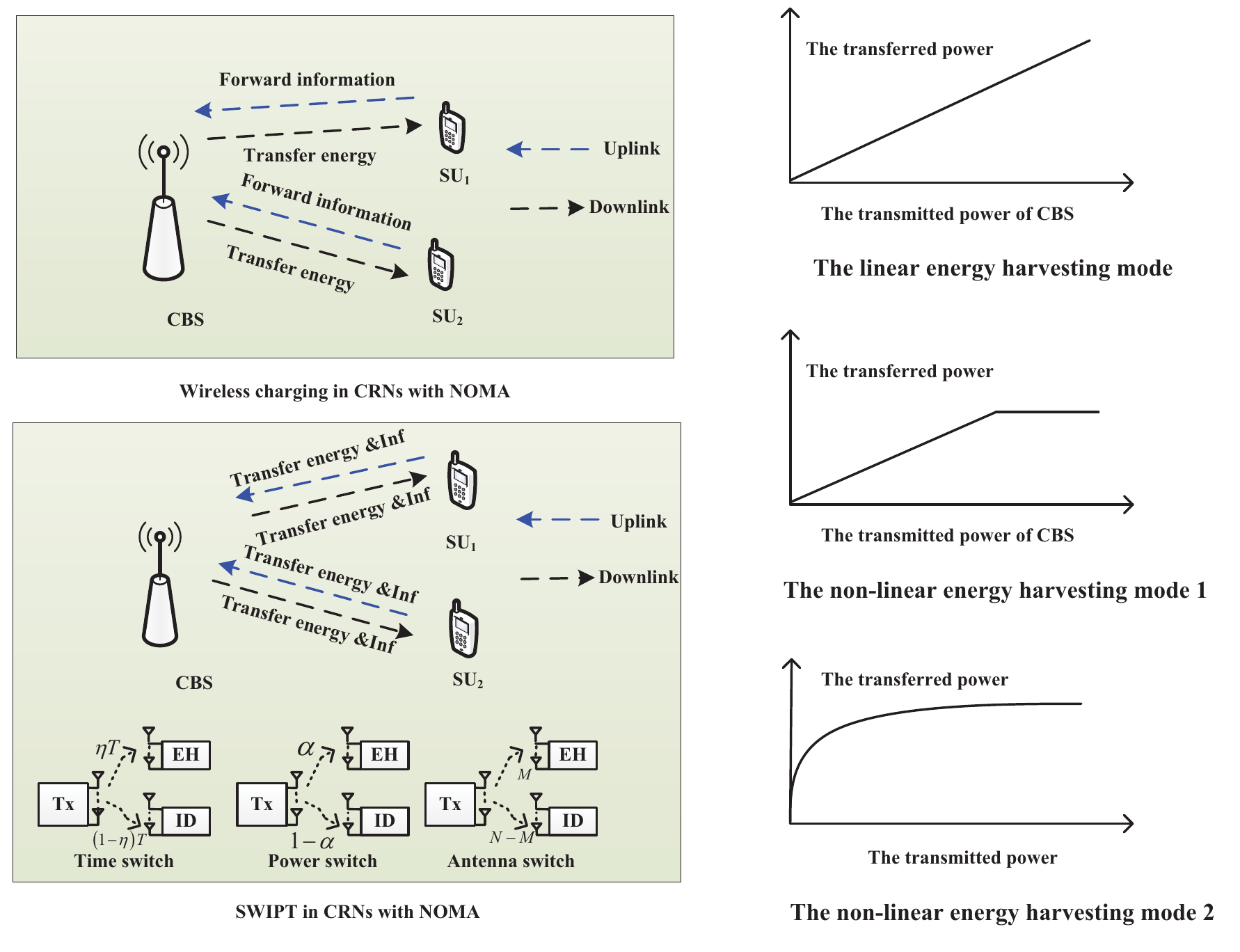}
\caption {\space\space Three different energy harvesting modes.}
\label{system model}
\end{figure*}
\textbf{Energy harvesting:} In CRNs with NOMA, there may exist energy-limited devices, such as sensors. EH techniques that leverage radio frequency signals to harvest energy are promising to extend the operation time of energy-limited devices. However, the energy harvesting efficiency (EHE) and secure issues are the two key challenges in CRNs with NOMA. The EHE depends on the EH circuit and the resource allocation strategy designed based on the EH model. Fig. 6 presents the existing three EH models, namely one linear EH model and two non-linear EH models \cite{E. Boshkovska}. The linear EH model is ideal and resource allocation strategies are relatively easy to design due to the simple EH model. The non-linear EH models are practical but it is challenging to design an optimal resource allocation strategy for maximizing the EHE due to the non-linear EH form, especially when the second non-linear EH model is applied. Besides the EHE, secure issues are extremely challenging in CRNs with NOMA. The reason is that malicious energy-harvesting SUs may disguise themselves as licensed NOMA SUs and enjoy the service provided by the CBS. How to leverage physical-layer security techniques to improve the security of CRNs with NOMA needs to be explored \cite{E. Boshkovska}.

\textbf{mmW communications:} Although integrating mmW communications into CRNs with NOMA has many exciting advantages, several obstacles are required to be addressed. When the primary network operates in the mmWave bands, the signals transmitted to PUs may suffer from the vulnerability of shadowing and the intermittent connectivity due to the ultra-high frequency. Thus, in order to protect the QoS of PUs, it is critical yet very challenging to design beamforming schemes \cite{Z. Ding}. Moreover, the existing investigations on mmW communications are mainly focused on point-to-point communications. How to design multiple users access mechanisms for mmW communications in CRNs with NOMA is another open issue since NOMA techniques are proposed to simultaneously serve for multiple SUs.

\textbf{Cooperative transmission:} In the overlay mode, NOMA SUs can provide cooperation for PUs in order to obtain the spectrum access opportunity \cite{S. M. R. Islam}. This cooperation is achieved by superimposing PUs’ signals onto SUs’s signals. In order to protect the QoS of PUs and improve the SE of CRNs with NOMA, it is crucial to select appropriate NOMA SUs to efficiently cooperate with PUs. Thus, how to design an optimal cooperative NOMA SUs scheduling scheme is important.

\textbf{OMA\&NOMA:} The combination of NOMA with OMA can efficiently manage the mutual interference between the primary and secondary network and improve SE of CRNs \cite{S. M. R. Islam}. However, the main challenge in achieving this hybrid multiple access is the implementation complexity, especially when there are large numbers of SUs. If all SUs simultaneously enjoy services in the same wideband by using NOMA techniques, the implementation complexity of SUs' receiver is extremely high due to the equipment of multi-user detection. In order to make full use of the advantage of the hybrid multiple access, user clustering techniques that pair SUs into multiple clusters need to be designed. The number of SUs in a cluster can be managed and SUs in a cluster can be served by using NOMA. Different clusters can be simultaneously served by using OMA.

\textbf{Resource allocation and optimization:} Resource allocation and optimization are vital in CRNs with NOMA. They aim for efficiently utilizing resource in terms of different objectives (e.g., SE and EE). They also play an important role in the interference management. Moreover, they have an effect on the feasibility of multiuser detection. For example, an optimal power allocation not only can efficiently improve EE but also protect the QoS of PUs \cite{Y. Zhang}. However, there are mainly two challenges to be addressed. One is that when there are multiple conflicting objectives required to be jointly optimized, such as EE and SE, how to design an optimal resource allocation scheme to achieve a better tradeoff among various conflicting objectives is extremely challenging. Although the multi-objective optimization can be exploited to achieve Pareto optimal solutions, the complexity of the designed algorithms may be very high. Moreover, since it is extremely difficult to obtain perfect CSI in CRNs with NOMA, the design of a robust resource allocation scheme is important yet very challenging. It needs to consider the constraints that can successfully perform  multi-user detection and the impact of the imperfect CSI on multi-user detection.
\section{Conclusion}
CRNs with NOMA have the potential to significantly improve the SE and increase the number of user connectivity. In order to enable NOMA techniques to be practically applied in CRNs, it is imperious to understand and tackle the challenges associated with it. For the purpose of understanding the challenges and realize the benefits of integrating NOMA techniques into CRNs, this article firstly provided a review of the state-of-the-art research efforts made to enable NOMA techniques to implement in CRNs. Moreover, we devised a thematic taxonomy to categorize and classify the literature. Furthermore, we outlined the key challenges to achieve CRNs with NOMA. Finally, several open research issues in CRNs with NOMA were presented as future research directions.

We concluded that the investigation on integrating NOMA techniques into CRNs is in its infancy. The application of NOMA techniques into CRNs has many attractive advantages, such as high SE and the ability of providing services for massive users. This application can be possible only when the presented challenges are addressed. Moreover, the discuss of the key challenges can help the domain researchers and designers to enable NOMA techniques for CRNs.

\end{document}